\documentclass[aps,showpacs,superscriptaddress,preprint]{revtex4}%
\usepackage{mathrsfs}
\usepackage{amsfonts}
\usepackage{amsmath}
\usepackage{amssymb}
\usepackage{graphicx}%
\begin{document}
\title{Probing crossover from analogous weak antilocalization to localization by an
Aharonov-Bohm interferometer on topological insulator surface}
\author{Zhen-Guo Fu}
\affiliation{State Key Laboratory for Superlattices and Microstructures, Institute of
Semiconductors, Chinese Academy of Sciences, P. O. Box 912, Beijing 100083,
People's Republic of China}
\affiliation{LCP, Institute of Applied Physics and Computational Mathematics, P.O. Box
8009, Beijing 100088, China}
\author{Ping Zhang}
\thanks{zhang\_ping@iapcm.ac.cn}
\affiliation{LCP, Institute of Applied Physics and Computational Mathematics, P.O. Box
8009, Beijing 100088, China}
\author{Shu-Shen Li}
\thanks{sslee@semi.ac.cn}
\affiliation{State Key Laboratory for Superlattices and Microstructures, Institute of
Semiconductors, Chinese Academy of Sciences, P. O. Box 912, Beijing 100083,
People's Republic of China}

\begin{abstract}
We propose a scanning tunneling microscopy Aharonov-Bohm (AB) interferometer
on the surface of a topological insulator (TI) to probe the crossover from
analogous weak antilocalization (WAL) to weak localization (WL) phenomenon via
the AB oscillations in spin-resolved local density of states (LDOS). Based on
our analytical and numerical results, we show that with increasing the energy
gap of TI surface states, the $\Phi_{0}/2$=$hc/2e$ periodic AB oscillations in
spin-resolved LDOS gradually transit into the $\Phi_{0}$ periodic oscillations.

\end{abstract}

\maketitle

Topological insulators (TI) have attracted substantial interest in the modern
condensed matter physics since their extraordinary edge and surface states
\cite{Hasan, Qi2011}. Following a series of theoretical predictions
\cite{Kane, Bernevig,Fu,Zhanghj}, a variety of two-dimensional \cite{Konig}
and three-dimensional \cite{Hsieh1,Chen,Xia, Hsieh3} TI materials have been
realized in recent experiments. The helical spin structure of Dirac electrons
in the gapless strong TIs acquire a spin-orbit induced nontrivial Berry phase
of $\pi$ after a $2\pi$ adiabatic rotation along the Fermi surface, which
results in prohibition of backscattering and the weak antilocalization (WAL).
The WAL effect in TIs has been measured by transport experiments
\cite{Olshanetsky,Ghaemi,He,Chen2,Liu1,Wang,Peng}. Through gradually doping Cr
magnetic elements in Bi$_{2}$Se$_{3}$ to open and increase the Dirac
electrons's energy gap, most recently, the crossover from WAL to weak
localization (WL) has been well observed \cite{Liu2}, consistent with
theoretical prediction \cite{Lu2011}.

Meanwhile, the surface scanning tunneling microscopy (STM) measurements have
been extensively carried out to study the electronic properties and impurity
scattering effects of TIs \cite{Hasan, Qi2011,Balatsky}. Compared to the
conventional transport techniques that give the averaged signals, the STM has
its own advantages in probing and even manipulating single-impurity and
multiple-impurity scattering, from which the more precise quantum processes
and mechanisms become possible to reveal. In particular, due to the nature of
the single Dirac cone, no complicated intervalley scattering events happen on
the TI surface, therefore, \textit{in situ} designing of specific impurity
configurations and mapping of their scattered electronic states can be
harnessed to image extraordinary quantum properties of TI surfaces. Recently,
we proposed a spin-dependent Aharonov-Bohm (AB) interferometer \cite{ZGFu},
which consists of a spin-polarized STM tip and two identical nonmagnetic
impurities, to probe the unique spin-texture related quantum scattering
behaviors on \textit{gapless} TI surfaces. Using this interferometer, we have
shown an interesting analogous WAL phenomenon reflected from $\Phi_{0}/2$
periodic AB oscillations in the spin-resolved local density of states (LDOS),
which is absent in the conventional metal surface systems.

\begin{figure}[ptb]
\begin{center}
\includegraphics[width=0.5\linewidth]{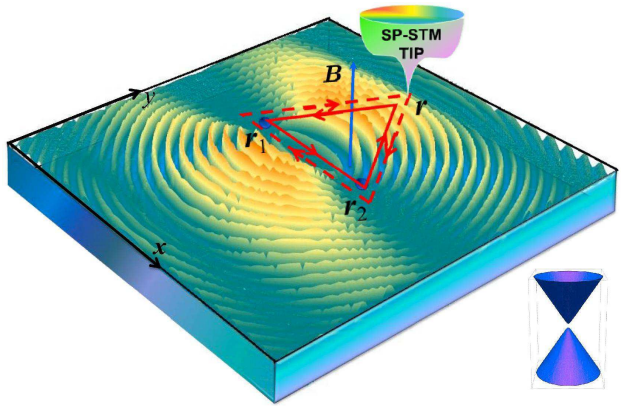}
\end{center}
\caption{ (Color online) Surface electronic interferometer, comprising a spin
polarized STM tip at $\boldsymbol{r}$ and two impurities at $\boldsymbol{r}%
_{1}$ and $\boldsymbol{r}_{2}$ separately. The interference contributions in
the LDOS is introduced by the electrons traveling along clockwise and
anticlockwise loops enclosed by the STM tip and two impurities. The applied
magnetic field $B$ affects this interference via the AB effect. }%
\label{fig1}%
\end{figure}

In the present paper, we extend to consider the spin-dependent AB effect on
\textit{gapped} TI surfaces. If the Dirac fermions gain mass by coupling to a
magnetic exchange field, a gap will be induced in the system, breaking time
reversal symmetry in scattering amplitudes, and as a consequence scattering
and transport properties for gapped and gapless systems should be
significantly different, as witnessed by the above-mentioned WAL to WL
crossover. Inspired by this idea, thus we study the effects of finite gap on
the AB oscillations in spin-resolved LDOS that we previously initiated. The AB
oscillations in the real-space LDOS maps are arisen from the magnetic flux
threaded through the time-reversed self-crossing loops (see Fig. \ref{fig1}).
We find on one hand that the total LDOS exhibits the AB oscillation with a
period of $\Phi_{0}$=$hc/e$ in both gapless and gapped cases. On the other
hand, while the spin-resolved LDOS shows $\Phi_{0}/2$ periodic AB oscillations
on the gapless TI surface, as we have reported before \cite{ZGFu}, however,
with increasing the energy gap, the $\Phi_{0}/2$ period of AB oscillations in
spin-resolved LDOS gradually disappears and at the same time the $\Phi_{0}$
period becomes clear. In other words, the crossover from analogous WAL to WL
on TI surface can be well observed using our spin-dependent AB interferometer
instead of the complicated low-temperature transport measurement. This
crossover is consistent with the evolution of the Berry phase with increasing
the TI surface-state gap.

The TI surface, on which two nonmagnetic impurities are adsorbed, is described
by a low-energy effective Dirac Hamiltonian
\begin{equation}
H=H_{0}+V\left(  \boldsymbol{r}\right)  , \label{e1}%
\end{equation}
where
\begin{equation}
H_{0}\mathtt{=}\hbar v_{f}\left(  k_{x}\sigma_{y}\mathtt{-}k_{y}\sigma
_{x}\right)  \mathtt{+}\Delta\sigma_{z},
\end{equation}
with $v_{f}$ ($\mathtt{\approx}4\mathtt{\sim}5\mathtt{\times}10^{5}$ m/s for
Bi$_{2}$Te$_{3}$-family TIs as numerically used throughout this paper) being
the Fermi velocity and $2\Delta$ the energy gap of massive Dirac fermions,
which is absent in the massless limit.
\begin{equation}
V(\boldsymbol{r})\mathtt{=}\sum_{i\mathtt{=}1}^{2}U_{i}\sigma_{0}\delta\left(
\boldsymbol{r}\mathtt{-}\boldsymbol{r}_{i}\right)
\end{equation}
denotes the potential of two impurities located at $\boldsymbol{r}%
_{1}\mathtt{=}\left(  \mathtt{-}d/2\mathtt{,}0\right)  $ and $\boldsymbol{r}%
_{2}\mathtt{=}\left(  d/2\mathtt{,}0\right)  $ with $U_{i}$ the potential
scattering strength for scalar impurities. $\sigma_{0}$ is the
$2\mathtt{\times}2$ unit matrix. 

The unperturbed real-space Green's function $G_{0}\left(  \boldsymbol{r}%
\mathtt{-}\boldsymbol{r}^{\prime},\omega\right)  $ can be obtained from the
Fourier transformation of $G_{0}\left(  \boldsymbol{k},i\omega\right)
$=$\left(  i\omega\text{\texttt{-}}H_{0}\right)  ^{-1}$ in the $k$-space,
which after a straightforward derivation is given by
\begin{align}
G_{0}\left(  \boldsymbol{r}\mathtt{-}\boldsymbol{r}^{\prime},\omega\right)
&  =\frac{\mathtt{-}i\omega}{\left(  2\hbar v_{f}\right)  ^{2}}\left[  \left(
1+\Delta/\omega\sigma_{z}\right)  H_{0}^{\left(  1\right)  }\left(  z\right)
\right. \nonumber\\
&  \left.  -i\left(  \boldsymbol{\hat{\rho}}\times\boldsymbol{\sigma}\right)
\cdot\boldsymbol{\hat{z}}\sqrt{\delta_{+}\delta_{-}}H_{1}^{\left(  1\right)
}\left(  z\right)  \right]  , \label{green}%
\end{align}
and
\begin{equation}
G_{0}\left(  \boldsymbol{0},\omega\right)  =\frac{\left\vert \omega\right\vert
}{2\pi}\text{\textrm{diag}}\left[  \delta_{+}f_{0},\delta_{-}f_{0}\right]
\end{equation}
for $\left\vert \omega\right\vert \mathtt{>}\Delta$, where $\delta_{\pm}%
$=$\left(  1\mathtt{\pm}\Delta/\omega\right)  $, $z\mathtt{=}\frac
{\sqrt{\delta_{+}\delta_{-}}\omega\rho}{\hbar v_{f}}$ and $f_{0}\mathtt{=}%
\int\frac{kdk}{\omega^{2}-\Delta^{2}-\left(  \hbar v_{f}k\right)  ^{2}}$.
Here, $\boldsymbol{\hat{\rho}}$ is the unit vector of $\boldsymbol{\rho
}\mathtt{=}\boldsymbol{r\mathtt{-}r}^{\prime}$ and $H_{0/1}^{\left(  1\right)
}\left(  z\right)  $ are the Hankel functions of the first kind.

The features we discuss are expected to be seen in the change of the
real-space LDOS owing to the influence of magnetic flux which passes through
the area enclosed by the two scattering paths shown in Fig. \ref{fig1}. This
quantity can reveal the analogous WL or WAL effect in TI via AB oscillatory
periods in LDOS. The real-space Green's function involving the impurities
scattering is given by Dyson equation $G\mathtt{=}G_{0}\mathtt{+}\delta G$,
with
\begin{equation}
\delta G=\int d\boldsymbol{r}^{\prime\prime}G_{0}\left(  \boldsymbol{r}%
-\boldsymbol{r}^{\prime\prime};\omega\right)  V\left(  \boldsymbol{r}%
^{\prime\prime}\right)  G\left(  \boldsymbol{r}^{\prime\prime},\boldsymbol{r}%
^{\prime};\omega\right)  . \label{e4}%
\end{equation}

Following the perturbation approach, Eq. (\ref{e4}) can be expanded to any
order in the impurity potential $V$. Our effort is concentrated on the
scattering processes of surface electrons with the both impurities, in which
the scattering paths enclose loops \cite{Cano}. Therefore, taking all this
into account, after a long algebra calculation, we have%
\begin{align}
\delta G_{L}  &  =G_{0}\left(  \boldsymbol{r}-\boldsymbol{r}_{1}\right)
W_{1}G_{0}\left(  \boldsymbol{r}_{1}-\boldsymbol{r}_{2}\right)  T_{2}%
G_{0}\left(  \boldsymbol{r}_{2}-\boldsymbol{r}^{\prime}\right) \nonumber\\
&  +G_{0}\left(  \boldsymbol{r}-\boldsymbol{r}_{2}\right)  W_{2}G_{0}\left(
\boldsymbol{r}_{2}-\boldsymbol{r}_{1}\right)  T_{1}G_{0}\left(  \boldsymbol{r}%
_{1}-\boldsymbol{r}^{\prime}\right)  , \label{loop}%
\end{align}
where the subscript $L$ represents the loops enclosed by the scattering paths
of the surface electrons, and
\begin{equation}
W_{i}=\frac{T_{i}}{\sigma_{0}-T_{i}G_{0}\left(  \boldsymbol{r}_{i}%
-\boldsymbol{r}_{j}\right)  T_{j}G_{0}\left(  \boldsymbol{r}_{i}%
-\boldsymbol{r}_{j}\right)  } \label{e6}%
\end{equation}
with $T\mathtt{-}$matrices $T_{i}\mathtt{=}\frac{V_{i}}{\sigma_{0}%
\mathtt{-}V_{i}G_{0}\left(  \mathbf{0}\mathtt{;}\omega\right)  }$ ($i,j$%
=$1,2$). Equation (\ref{loop}) is a general formula describing the
two-impurity back and forth scattering of the STM-probed quasiparticles. Thus,
the interference information during the time-reversal scattering processes are
included in this equation.

In the presence of a weak magnetic field, the Green's function can be
semiclassically approximated as
\begin{equation}
\widetilde{G}_{0}\left(  \boldsymbol{r}-\boldsymbol{r}^{\prime},\omega\right)
=e^{i\frac{2\pi}{\Phi_{0}}\int_{\boldsymbol{r}}^{\boldsymbol{r}^{\prime}%
}\boldsymbol{A}\left(  \boldsymbol{l}\right)  \cdot d\boldsymbol{l}}%
G_{0}\left(  \boldsymbol{r}-\boldsymbol{r}^{\prime},\omega\right)  ,
\label{app}%
\end{equation}
where $\boldsymbol{A}\mathtt{=}\left(  \mathtt{-}By,0,0\right)  $ represents
the vector potential. The correction of the LDOS due to the magnetic flux is
given by%
\begin{align}
\Delta N_{L}\left(  \boldsymbol{r},\omega,B\right)   &  =-\frac{1}{\pi
}\operatorname{Im}\text{Tr}\left[  \delta\widetilde{G}_{L}\left(
\boldsymbol{r},\omega\right)  -\delta G_{L}\left(  \boldsymbol{r}%
,\omega\right)  \right] \nonumber\\
&  =\Delta N_{L}^{\uparrow}\left(  \boldsymbol{r},\omega,B\right)  +\Delta
N_{L}^{\downarrow}\left(  \boldsymbol{r},\omega,B\right)  , \label{LDOS}%
\end{align}
where $\delta\widetilde{G}_{L}$ is calculated from Eq. (\ref{loop}) with
$\widetilde{G}_{0}$. It is clear that the magnetic field affects the LDOS via
the magnetic flux, which is easy to be obtained from the integral over the
loops ($\boldsymbol{r}\mathtt{\rightleftharpoons}\boldsymbol{r}_{1}%
\mathtt{\rightleftharpoons}\boldsymbol{r}_{2}\mathtt{\rightleftharpoons
}\boldsymbol{r}$), $\oint\nolimits_{\boldsymbol{l}}\boldsymbol{A}\left(
\boldsymbol{l}\right)  \mathtt{\cdot}d\boldsymbol{l}\mathtt{=}\mathtt{\pm
}Bdy/2\mathtt{=}\mathtt{\pm}\Phi$. In the present setup, we focus solely on
the (spin-resolved) LDOS at $\boldsymbol{r}\mathtt{=}(x,y)$, which is probed
by the STM tip with the same plane coordinates. Since the STM tip also dually
participate in composing the closed trajectory that the Dirac electron
travels, hence under a fixed magnetic field $B$, the AB interference displays
the oscillations with varying the tip position along the $y$ direction.

The condition $l_{B}\mathtt{>}\lambda_{f}$ of the semiclassical approximation
in Eq. (\ref{app}) should be satisfied, here $B\mathtt{=}5$ T and
$\varepsilon_{f}\mathtt{=}150$ meV are chosen from which the corresponding
magnetic length $l_{B}\mathtt{\approx}11.63$ nm while the Fermi wave length
$\lambda_{f}\mathtt{=}10.5$ nm. Also, the Zeeman splitting by the external
magnetic field is negligibly small (typically of $0.5$ meV at $B\mathtt{=}5$ T
for Bi$_{2}$Se$_{3}$ film \cite{WangZG}) compared to the strong spin-orbit
coupling (SOC), and thereby is neglected in this discussion. In a recent low
temperature transport experiment \cite{Liu2} on magnetically doped Bi$_{2}%
$Se$_{3}$ film, an energy gap as large as $100\mathtt{\sim}300$ meV near the
Dirac point was observed, which is comparative with our choise in this work.

By choosing suitable Fermi energy $\varepsilon_{f}$ and energy gap parameter
$\Delta$, which can be controlled in experiments, the backscattering and the
crossover from the WAL to WL can occur on TI surface. When the Fermi energy
lies in the gap, the interference signals arising from the contributions of
$\Delta N_{L}\left(  \boldsymbol{r},\omega,B\text{=}0\right)  $ are so weak
that the oscillatory ellipse features as well as the AB effect become
ambiguous, hence we only consider $\varepsilon_{f}$%
%TCIMACRO{\TEXTsymbol{>}}%
%BeginExpansion
$>$%
%EndExpansion
$\Delta$.

When the energy gap is opened, differing from the gapless case, the
$T\mathtt{-}$matrices $T_{i}\mathtt{=}\frac{V_{i}}{\sigma_{0}\mathtt{-}%
V_{i}G_{0}\left(  \mathbf{0}\mathtt{;}\omega\right)  }$ ($i,j$=$1,2$) are not
proportional to a unit matrix because $G_{0}^{11}\left(  \mathbf{0}%
\mathtt{;}\omega\right)  \mathtt{\neq}G_{0}^{22}\left(  \mathbf{0}%
\mathtt{;}\omega\right)  $, thus the $W_{i}$ matrices in Eq. (\ref{e6}) are no
longer diagonal, resulting in intractable complexity in reducing Eq. (10). To
get the asymptotic expression of LDOS we consider the lowest order in the
impurity potential in Eq. (\ref{loop}), which is given by%
\begin{align}
\delta G_{L}^{\left(  2\right)  }  &  =U^{2}G_{0}\left(  \boldsymbol{r}%
-\boldsymbol{r}_{1}\right)  G_{0}\left(  \boldsymbol{r}_{1}-\boldsymbol{r}%
_{2}\right)  G_{0}\left(  \boldsymbol{r}_{2}-\boldsymbol{r}^{\prime}\right)
\nonumber\\
&  +U^{2}G_{0}\left(  \boldsymbol{r}-\boldsymbol{r}_{2}\right)  G_{0}\left(
\boldsymbol{r}_{2}-\boldsymbol{r}_{1}\right)  G_{0}\left(  \boldsymbol{r}%
_{1}-\boldsymbol{r}^{\prime}\right)  . \label{ap1}%
\end{align}
For large distances ($\omega\rho/\hbar v_{f}\mathtt{\gg}1$), the Hankel
functions can be approximated as $H_{0/1}^{\left(  1\right)  }\left(
z\right)  \mathtt{\approx}\mathtt{\pm}\sqrt{\frac{2}{\pi z}}e^{i\left(
z\mp\pi/4\right)  }$, then the unperturbed Green's function has a simple
asymptotic form
\begin{equation}
G_{0}\left(  \mathbf{r}-\mathbf{r}^{\prime}\right)  \approx\frac
{-i\sqrt{\omega/2\pi\rho\hbar v_{f}}}{2\hbar v_{f}}\left(
\begin{array}
[c]{cc}%
\delta_{+}e^{i\left(  z-\frac{\pi}{4}\right)  } & \sqrt{\delta_{+}\delta_{-}%
}e^{-i\vartheta}e^{i\left(  z+\frac{\pi}{4}\right)  }\\
-\sqrt{\delta_{+}\delta_{-}}e^{i\vartheta}e^{i\left(  z+\frac{\pi}{4}\right)
} & \delta_{-}e^{i\left(  z-\frac{\pi}{4}\right)  }%
\end{array}
\right)  , \label{ap2}%
\end{equation}
where $e^{i\vartheta}\mathtt{=}\frac{\boldsymbol{\rho}\cdot\left(  \hat
{x}\mathtt{+}i\hat{y}\right)  }{\rho}$. Substituting this equation into Eq.
(\ref{ap1}) and after a tedious derivation, we obtain an explicit expression
of the total LDOS as follows%
\begin{align}
\Delta N_{L}^{\left(  2\right)  }\left(  \boldsymbol{r},\omega,B\right)   &
\approx C\sin\left(  \frac{\pi\Phi}{\Phi_{0}}\right)  \left\{  \left(
\delta_{+}^{3}+\delta_{-}^{3}\right)  \sin\left(  \frac{\pi\Phi}{\Phi_{0}%
}\right)  \right. \nonumber\\
&  -\delta_{+}^{2}\delta_{-}\left[  \sin\left(  \frac{\pi\Phi}{\Phi_{0}%
}-\vartheta_{1}\right)  +\sin\left(  \frac{\pi\Phi}{\Phi_{0}}-\vartheta
_{2}\right)  +\sin\left(  \frac{\pi\Phi}{\Phi_{0}}+2\phi\right)  \right]
\nonumber\\
&  \left.  -\delta_{-}^{2}\delta_{+}\left[  \sin\left(  \frac{\pi\Phi}%
{\Phi_{0}}+\vartheta_{1}\right)  -\sin\left(  \frac{\pi\Phi}{\Phi_{0}%
}-\vartheta_{2}\right)  +\sin\left(  \frac{\pi\Phi}{\Phi_{0}}-2\phi\right)
\right]  \right\}  ,
\end{align}
where $C\mathtt{=}4U^{2}\left(  \omega/8\pi(\hbar v_{f})^{3}\right)
^{3/2}\left(  1/\rho_{1}\rho_{2}d\right)  ^{1/2}\cos\left(  \frac{\sqrt
{\delta_{+}\delta_{-}}\omega\left(  \rho_{1}+\rho_{2}+d\right)  }{\hbar v_{f}%
}\mathtt{-}\frac{\pi}{4}\right)  $, $\phi\mathtt{=}\frac{\vartheta
_{1}\mathtt{-}\vartheta_{2}}{2}$ with $e^{i\vartheta_{1/2}}\mathtt{=}%
\frac{\boldsymbol{\rho}_{1/2}\mathtt{\cdot}(\hat{x}\mathtt{+}i\hat{y})}%
{\rho_{1/2}}$ and $\boldsymbol{\rho}_{1/2}\mathtt{=}\boldsymbol{r}%
\mathtt{-}\boldsymbol{r}_{1/2}$. The spin-up and spin-down LDOSs are written
as
\begin{align}
\Delta N_{L}^{\left(  2\right)  \uparrow/\downarrow}\left(  \boldsymbol{r}%
,\omega,B\right)   &  \approx C\delta_{\pm}\sin\left(  \frac{\pi\Phi}{\Phi
_{0}}\right)  \left\{  \delta_{\pm}^{2}\sin\left(  \frac{\pi\Phi}{\Phi_{0}%
}\right)  -\delta_{\mp}^{2}\sin\left(  \frac{\pi\Phi}{\Phi_{0}}\mp
2\phi\right)  \right. \nonumber\\
&  +\left.  \delta_{+}\delta_{-}\left[  \sin\left(  \frac{\pi\Phi}{\Phi_{0}%
}\pm\vartheta_{2}\right)  -\sin\left(  \frac{\pi\Phi}{\Phi_{0}}\mp
\vartheta_{1}\right)  \right]  \right\}  .
\end{align}
Equations (13) and (14) are the main analytical result of this paper. Although
the total and spin-resolved LDOSs are now explicitly obtained, their AB
oscillation periods are still difficult to analytically determine due to the
weight coefficients $\delta_{+}$ and $\delta_{-}$ that arise from the energy
gap. However, there exist two extreme cases: (i) One is the gapless limit
where $\delta_{+}$=$\delta_{-}$=$1$; (ii) The other one is the large-gap
limit. For these two cases, one can further analytically simplify Eqs. (13)
and (14) and get the periods of real-space AB oscillations in the LDOSs of TI.

The gapless case have been discussed in our last paper \cite{ZGFu}. Actually,
when $\Delta$=$0$ we have the total LDOS quantity
\begin{equation}
\left.  \Delta N_{L}^{\left(  2\right)  }\left(  \boldsymbol{r},\omega
,B\right)  \right\vert _{\Delta=0}\varpropto\left[  \cos\left(  2\pi\Phi
/\Phi_{0}\right)  -1\right]  \label{e9}%
\end{equation}
and its two spin compents
\begin{equation}
\left.  \Delta N_{L}^{\left(  2\right)  \uparrow/\downarrow}\left(
\boldsymbol{r},\omega,B\right)  \right\vert _{\Delta=0}\varpropto\left[
\sin\left(  2\pi\Phi/\Phi_{0}\mp\phi\right)  \pm\sin\phi\right]  .
\label{spin}%
\end{equation}
Obviously, the spacial AB oscillation period\ of the total LDOS is $y_{0}%
$=$\frac{2\Phi_{0}}{Bd}$ for fixed $B$ and fixed impurity configuration, which
corresponds to a $\Phi_{0}$ period in the scale of flux. \begin{figure}[ptb]
\begin{center}
\includegraphics[width=0.6\linewidth]{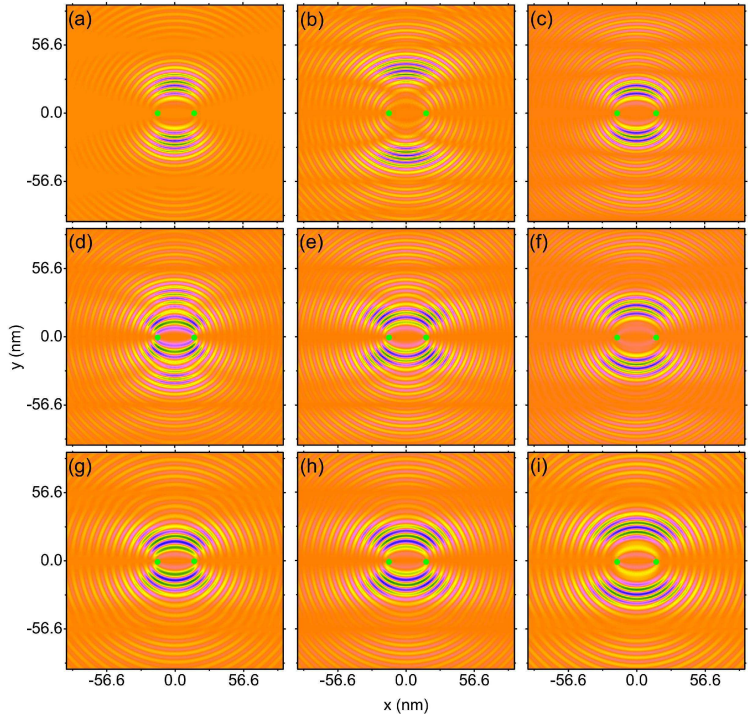}
\end{center}
\caption{ (Color online) Simulations of the AB oscillations of the electronic
LDOS in Bi$_{2}$Te$_{3}$(111) surface with two nonmagnetic impurities under a
magnetic field $B\mathtt{=}5$ T. The left, middle, and right panels correspond
to the total, spin-up, and spin-down LDOSs, respectively. We choose
$\Delta/\varepsilon_{f}\mathtt{=}0$ in (a-c), 0.3 in (d-f), and 0.6 in (g-i)
with $\varepsilon_{f}\mathtt{=}150$ meV. The horizonal strips in patterns are
signature of AB effect. The green dots denote the impurity positions. The
other parameters are chosen as $v_{f}\mathtt{=}4\mathtt{\times}10^{5}$ m$/$s,
$d\mathtt{=}30$ nm, and $U\mathtt{=}1$ eV.}%
\label{fig2}%
\end{figure}

As clearly seen from Eq. (\ref{spin}), comparing to the total LDOS, there
occurs in the spin-resolved LDOS additional strong SOC induced quantum
interference signature. This strong spin interference effect deviates the
real-space AB oscillations from $y_{0}$ period when a spin-polarized STM tip
scans along the $\hat{y}$-direction on the TI surface in the presence of a
fixed $B$. The AB interference period along the $\hat{y}$-direction can be
numerically determined by solving the zero-point equation of the factors in
Eq. (\ref{spin}), i.e., $\sin(\frac{2\pi\Phi}{\Phi_{0}}\mathtt{\mp}%
\phi)\mathtt{\pm}\sin\phi$=$0$. We can get the asymptotic roots
$y\mathtt{\approx}\frac{n\Phi_{0}}{Bd}$ ($n\mathtt{\in}\mathbb{Z}$) for weak
$B$. Therefore, the AB oscillation signals for $\left.  \Delta N_{L}^{\left(
2\right)  \uparrow/\downarrow}\right\vert _{\Delta=0}$ occur at $\mathtt{\sim
}\frac{n\Phi_{0}}{Bd}$ with a spacial period of $\frac{\Phi_{0}}{Bd}%
\mathtt{=}\frac{a_{0}}{2}$ (i.e., $\frac{\Phi_{0}}{2}$ in the scale of flux),
the half of $\Delta N_{L}$. This half period could be understood as an analog
of WAL effect in the spin-resolved LDOS.

Turning to the other extreme case that the gap is large enough to be
comparable with the Fermi energy $\varepsilon_{f}$. In this case, Eq. (14) is
simplified to be
\begin{align}
\Delta N_{L}^{\left(  2\right)  \uparrow}\left(  \boldsymbol{r},\omega
,B\right)   &  \propto\delta_{+}^{3}\left[  \cos\left(  2\pi\Phi/\Phi
_{0}\right)  -1\right] \nonumber\\
\Delta N_{L}^{\left(  2\right)  \downarrow}\left(  \boldsymbol{r}%
,\omega,B\right)   &  \propto\delta_{+}^{2}\delta_{-}\left[  \cos\left(
2\pi\Phi/\Phi_{0}\right)  -1\right]
\end{align}
near the Fermi energy. Clearly, the spin-resolved LDOS $\Delta N_{L}^{\left(
2\right)  \uparrow/\downarrow}$ turns now to display a complete $\Phi_{0}$
period of AB oscillations as the total LDOS $\Delta N_{L}^{\left(  2\right)
}$ does, which is totally different from the gapless case. In other words, the
spin-resolved LDOS clearly exhibits an analogous WL phenomenon in the
large-gap limit. Except for these two extreme limits, to observe the AB
oscillation period of spin-resolved LDOS for intermediate values of the energy
gap, we resort to exact numerical analysis on original Eqs. (\ref{green}%
-\ref{LDOS}).

From numerical calculations, we find by tuning the ratio of $\Delta
/\varepsilon_{f}$ that the crossover from half-period $\Phi_{0}/2$ to period
$\Phi_{0}$ emerges in the AB oscillations of the spin-resolved LDOS. When
$\Delta/\varepsilon_{f}$ is small, the spacial AB oscillations of the
spin-resolved LDOS still approximately possess a half period of $\Phi_{0}/2$,
i.e., the analogous WAL effect is dominant in spin-resolved LDOS. However,
with increasing $\Delta/\varepsilon_{f}$, the $\Phi_{0}/2$ period of AB effect
gradually disappears and at the same time the $\Phi_{0}$ period becomes
obvious in spin-resolved LDOS.

The typical numerical results are shown in Fig. \ref{fig2} for different
values of $\Delta/\varepsilon_{f}$. The horizontal strips in each panel are
the AB oscillation signals in the real-space LDOS. It is obvious that in the
case of $\Delta$=$0$, the interstrip distance is $y_{0}\mathtt{=}\frac
{2\Phi_{0}}{Bd}$=$56.6$ nm in the total LDOS as shown in Fig. \ref{fig2}(a),
corresponding to the $\Phi_{0}$ period of AB oscillations in total LDOS.
Whereas, the interstrip distance becomes $y_{0}/2$=$28.3$ nm in the
spin-resolved LDOS as shown in Figs. \ref{fig2}(b) and \ref{fig2}(c),
corresponding to the $\frac{\Phi_{0}}{2}$ period of AB oscillations, which is
consistent with the above analysis on Eq. (\ref{spin}). From Figs.
\ref{fig2}(d-f) corresponding to $\Delta/\varepsilon_{f}$=0.3, we can observe
that the horizontal strips in spin-resolved LDOSs patterns move towards the
horizontal strips in total LDOS, signifying a crossover from analogous WAL to
WL in the spin-resolved LDOS. With further increasing the ratio of
$\Delta/\varepsilon_{f}$, the spin-resolved LDOS take on $\Phi_{0}$ periodic
AB oscillations, see the downmost panels (g-i) in Fig. \ref{fig2} where
$\Delta/\varepsilon_{f}$ is chosen to be as large as 0.6. Much smaller values
of $B$ have also been tested in simulations, and the calculated AB oscillation
patterns of (spin-resolved) LDOS are similar to those shown herein. Thus that,
the AB effect can be effectively studied in a semiclassical way in a wide
range of $B$.

Our findings can be understood from the view point of Berry phase. The TI
surface Dirac electrons traveling along two time-reversed self-crossing loops
($\boldsymbol{r}\mathtt{\rightleftharpoons}\boldsymbol{r}_{1}%
\mathtt{\rightleftharpoons}\boldsymbol{r}_{2}\mathtt{\rightleftharpoons
}\boldsymbol{r}$) differentiate by a Berry phase associated with spin rotation
of $2\pi$, which is given by $-i\int_{0}^{2\pi}d\theta_{k}\left\langle
\psi_{k}\right\vert \partial_{\theta_{k}}\left\vert \psi_{k}\right\rangle
\mathtt{=}(1\mathtt{-}\Delta/\varepsilon)\pi$. Here, the eigenstates of
$H_{0}$ are expressed as $\left\vert \psi_{k}\right\rangle $=$\frac
{e^{i\boldsymbol{k}\cdot\boldsymbol{r}}}{\sqrt{2}}\left(
\begin{array}
[c]{cc}%
\gamma_{+}, & \mp i\gamma_{-}e^{i\theta_{k}}%
\end{array}
\right)  ^{T}$, where $\gamma_{\pm}$=$\sqrt{1\mathtt{\pm}\Delta/\varepsilon}$,
$\tan\theta_{k}\mathtt{=}k_{y}/k_{x}$, and $\varepsilon\mathtt{=}\sqrt{\left(
\hbar v_{F}k\right)  ^{2}\mathtt{+}\Delta^{2}}$. The two clockwise and
anticlockwise loops enclosed by the STM tip and two impurities in Fig. 1
accumulate a Berry phase of $\pi$ on a gapless TI surface due to the
spin-momentum locking, and result in the WAL effect, which are represented as
AB oscillations with half-period of $\Phi_{0}/2$ in the spin-resolved LDOSs.
While with the opening of the gap, the Berry phase departures from $\pi$. The
larger ratio of $\Delta/\varepsilon_{f}$ causes greater deviation of the Berry
phase from $\pi$, leading to stronger WL tendency which is consistent with the
observations from the AB interferometer proposed here. Therefore, the
spin-dependent AB interferometer shown in this paper may provide a feasible
approach to study the competition between WAL and WL by observing the spatial
AB oscillation periods in the spin-resolved LDOS maps.

To experimentally verify our predictions exhibited here, the spin polarized
STM technique is required, which we believe is achievable
\cite{Bergmann1,Schmaus}. Finally, we should point out that dephasing
processes have been observed in transport investigations in Bi$_{2}$Se$_{3}$
and Bi$_{2}$Te$_{3}$ films \cite{Wang, He,Liu2} as well as in AB-effect
studies of Bi$_{2}$Se$_{3}$ nanowires \cite{Peng}. The phase coherence length
$l_{\phi}$ of Bi$_{2}$Se$_{3}$ and Bi$_{2}$Te$_{3}$ can be as large as
hundreds of nanometers, which is tens times of the Fermi wave length. The
characteristic distance in our setup must be much smaller than the phase
coherence length ($d\mathtt{\ll}l_{\phi}$), so that it is reasonable in Fig. 2
to choose $d\mathtt{=}30$nm$\mathtt{\ll}l_{\phi}$ without taking into account
the dephasing processes in the above numerical calculations.

In summary, we have performed a semiclassical analysis of the spin polarized
STM probed AB oscillations in the LDOS induced by two impurities on a TI
surface. With increasing the surface gap of TI, the crossover from analogous
WAL to WL has been found in the AB oscillations of spin-resolved LDOS. This
phenomenon may provide an important alternative approach to testify various
extraordinary quantum wavefunction properties on the TI surface.

This work was supported by NSFC under Grants No. 90921003 and No. 60821061,
and by the National Basic Research Program of China (973 Program) under Grants
No. 2009CB929103 and No. G2009CB929300.

\end{document}